\documentclass[12pt]{iopart}
\usepackage{graphicx}
\usepackage{subfigure}
\usepackage{iopams}
\expandafter\let\csname equation*\endcsname\relax
\expandafter\let\csname endequation*\endcsname\relax
\usepackage{amsmath}

\renewcommand{\Im}{\mathop{\rm Im}\nolimits}
\renewcommand{\Re}{\mathop{\rm Re}\nolimits}
\renewcommand{\thesection}{\Roman{section}}

\begin{document}
\title[Optimal Stochastic Regularization]{Analytic Continuation of Quantum Monte Carlo Data: Optimal Stochastic Regularization Approach}
\author{I. S. Krivenko$^{1,2}$ and A. N. Rubtsov$^1$}

\address{$^1$ Department of Physics, Moscow State University, 119992 Moscow, Russia}
\address{$^2$ Centre de Physique Th\'{e}orique, Ecole Polytechnique, CNRS, 91128 Palaiseau Cedex, France}
\ead{igor@shg.ru}

\date{\today}

\begin{abstract}
A new algorithm for analytic continuation of noisy quantum Monte Carlo
(QMC) data from the Matsubara domain to real frequencies is proposed.
Unlike the widely used maximum-entropy (MaxEnt) procedure, our method
is linear with respect to input data and can therefore be applied to 
off-diagonal components of a thermal Green's function, or to a self-energy function.
The latter possibility is used to analyze QMC results for the half-filled
single-band Hubbard model on a Bethe lattice at a low temperature. Our method qualitatively
resolves peaks near the inner edges of the Hubbard bands in
the vicinity of a Mott transition, whereas a MaxEnt procedure does not.
An existence of such structures has been clearly established before in a high-precision
D-DMRG calculation by Karski \textit{et al}. We also analyze a stability of
the new method subject to changes of adjustable parameters.
\end{abstract}

\submitto{\JPCM}
\maketitle

\section{Introduction}
Modern theoretical analysis of strongly correlated systems
deals with a wide range of numerical methods, because no serious
progress has been made in formulating a regular analytic description yet.
In particular, a considerable progress has been achieved with the 
Dynamical Mean Field Theory (DMFT) \cite{RevModPhys.68.13,kotliar:865},
as well as with its extensions and generalizations \cite{maier:1027}.
In the DMFT approach a strongly-correlated lattice is reduced to an effective
impurity model. Virtually, the lattice is replaced with a single atom or
with a small cluster put into a Gaussian bath whose properties are defined in a self-consistent way.
DMFT requires a so-called solver algorithm, which is intended to give an approximate
evaluation of Green's functions for the effective impurity model.
For most cases, the Quantum Monte Carlo (QMC) class of solvers is used
\cite{PhysRevLett.56.2521}, \cite{PhysRevB.40.4780}.
These algorithms allow to calculate an electronic Green's function for many models
far beyond the regime where a perturbation theory is valid.
It is important to note that output of a QMC calculation belongs to the imaginary-time domain,
since DMFT equations at a finite temperature are written in terms of fermionic
Matsubara frequencies $\omega_n = (2n+1)\pi/\beta$, where $\beta$ is
an inverse temperature.

On the other hand, experimental spectral data are obtained in the real-frequency domain.
Consequently, it is evidently quite a difficult problem
to extract valuable physical information from a numerical data set, because
analytic continuation is required.

In the canonical formulation of the problem one has to obtain a spectral density
function $A(\omega)$ (we also use the term ``density of states'' as a synonym through this paper)
from a noisy set of values for a thermal Green's function $G(\tau)$ or equivalently for its Fourier transform
$G(i\omega_n) = \int_0^\beta d\tau\
e^{-i\omega_n\tau}G(\tau)$, produced in a QMC simulation. 
The spectral density is defined to be proportional to the imaginary part of a retarded Green's function:
\begin{equation*}
    A(\omega) = -\frac{1}{\pi}\Im G^R(\omega)
\end{equation*}
 
Due to analyticity of the retarded Green's function in the upper complex half-plane of frequency
the following integral representation holds:
\begin{equation}\label{Analyticity}
    G^R(\omega) = -\frac{1}{\pi}\int\limits_{-\infty}^{+\infty}\frac{\Im G^R(\omega')}{\omega - \omega' + i0}\ d\omega' =
    \int\limits_{-\infty}^{+\infty}\frac{A(\omega')}{\omega - \omega' +i0} d\omega'
\end{equation}
 
Doing a Fourier transform and substituting $\omega = i\omega_n$ one can derive the integral equation
\begin{equation}\label{MaxEnt_problem}
    G(\tau) = \int\limits_{-\infty}^{+\infty}d\omega
    \frac{e^{-\tau\omega}}{1 + e^{-\beta\omega}} A(\omega)
\end{equation}
which is a particular formulation of the analytic continuation
problem. The kernel of this integral equation is exponentially
small at large positive and negative frequencies, therefore a tiny
variation in $G(\tau)$ may correspond to a very strong
change of the spectrum at those frequencies. An uncertainty in
$G(\tau)$ is unavoidable due to the stochastic nature of
QMC algorithms, so the problem of numerical analytic continuation
is extremely ill-posed, similarly to a numerical inversion of the
Laplace transform.
 
The earliest attempts to solve the problem were based on the Pade
approximation method \cite{thirumalai:5029}. For a lot of cases,
modifications of the standard least squares procedure have been
proposed to solve integral equation (\ref{MaxEnt_problem})
\cite{PhysRevLett.55.1204,PhysRevLett.63.2504}.
Currently the most common practice is to use a Maximum Entropy
algorithm for the analytic continuation problem \cite{MaxEnt_Rev}.
It's based on a Bayesian inference concept and essentially uses
some \textit{a priori} knowledge about properties of the spectral
function, namely its positivity and the sum rule.
There are also the stochastic analytical inference method \cite{PhysRevE.81.056701}
and the method proposed by Mishchenko \textit{et al} \cite{PhysRevB.62.6317}.
Both of them rely on the same \textit{a priori} assumptions.
Taking into account these assumptions means that a MaxEnt algorithm
is essentially nonlinear in input data.
Existing implementations of MaxEnt do not allow the analytic
continuation of functions, whose norm is not a known constant,
or do not have a definite sign (a self-energy function or 
off-diagonal components of a Green's function, for instance).
 
The aim of this paper is to introduce a new analytic continuation
algorithm, which is linear with respect to input data and reliable
enough to be considered as an alternative to the MaxEnt method.
Section II of the paper is devoted to the general idea behind the
method without giving details of a particular implementation. It
explains the optimal regularization principle which is based on
the Tikhonov regularization ansatz. Some more specific implementation
details are described in sections III and IV. They in principle
may be revised and adjusted to some extent for a specific problem.
In Section V we present practical results of analytic
continuation for the single-band Hubbard model on a Bethe lattice,
which are then compared with corresponding results from the MaxEnt
algorithm. A brief stability analysis of the method is also presented
in that section. In Section VI we conclude the paper.
 
\section{The optimal regularization functional}

\newcommand{\bfx}{\mathbf{x}}
\newcommand{\bfy}{\mathbf{y}}
 
Generally speaking, the analytic continuation requires a solution of
a linear equation for an unknown complex function $F(\omega)$
\begin{equation}\label{REG_problem}
    \hat M F(\omega) = F(i\omega)
\end{equation}
using some \textit{a priori} information. Equation (\ref{MaxEnt_problem})
is a reformulated, but equivalent problem. Here $\hat M$ is an
ill-posed linear integral operator that analytically continues from the
real to the imaginary axis, and function $F(i\omega)$ is
approximately known from a QMC simulation.
 
For the case of Green's functions $F(\omega) \equiv
G^R(\omega)$ and $F(i\omega) \equiv G(i\omega_n)$ (some
QMC algorithms can provide $G(\tau)$ as well as its
Fourier tranform $G(i\omega_n)$).
 
If we introduce certain bases of orthogonal functions to
represent $F(\omega)$ and $F(i\omega)$, the above integral
equation will become a system of linear algebraic equations
\begin{equation}\label{first_eq}
    M \bfx=\bfy,
\end{equation}
where $M$ is a badly conditioned matrix, and the right hand part
$\bfy$ is defined with a certain level of inaccuracy. Formally, the system includes
an infinite number of equations, but one would expect that for a properly
chosen basis an effective reduction to a finite system is possible.
Such a basis, suitable for the analytic continuation problem is introduced in Section III.

The staring point of the proposed method is a Tikhonov
regularization of the problem \cite{Tikhonov}. Let us search for a
vector $\bfx$, which minimizes the Tikhonov functional:
\begin{equation}\label{tikhonov}
    \mathcal{F}[\bfx; R] = ||M \bfx - \bfy||^2 +(\bfx,R \bfx)
\end{equation}
Here $R$ is a regularizing Hermitian matrix. Vector $\mathbf{y}$ is
known approximately: $\mathbf{y} = \bar \bfy + \delta \bfy$, where $\bar \bfy$ is
a mean value of the vector $\bfy$, and a deviation $\delta \bfy$ is
a random quantity distributed with a zero mean value and
characterized by a covariation matrix $\hat K_{\bfy}$, such that:
\begin{equation}\label{K_def}
    \overline{\delta\bfy} = 0, \quad \hat K_{\bfy} = \overline{\delta\bfy \delta \bfy^\dag}
\end{equation}
(hereafter a line over an expression denotes the QMC expectation values).

Varying the functional $\mathcal{F}[\bfx;R]$ with respect to
$\bfx$ one obtains a condition for the vector $\bfx$, which gives a
minimum of the functional for a given $\bfy$ and $R$:
\begin{equation}\label{X_def}
    \bfx = X M^\dag \bfy, \quad \textrm{where} \ X \equiv
    (M^\dag M + R)^{-1}
\end{equation}

Let us assume that the vector $\bar{\bfx}$ is an exact solution of system 
(\ref{first_eq}) with a precisely known right-hand
part: $M\bar \bfx = \bar \bfy$. We average the mean square
deviation of $\bfx$ from $\bar \bfx$ over all possible values of the
random vector $\delta \bfy$ to obtain the following:
\begin{equation}\label{deviation}
    \overline{||\bfx - \bar \bfx||^2} = \mathrm{Tr}\{X A X -
    2 X B\} + \mathrm{Tr}\{\bar \bfx \bar \bfx^\dag\}
    \end{equation}
    \begin{equation}\label{A_def}
    A \equiv M^\dag M\bar \bfx\bar \bfx^\dag M^\dag M
    + M^\dag K_{\bfy} M
\end{equation}
\begin{equation}\label{B_def}
    B \equiv M^\dag M \bar \bfx \bar \bfx^\dag
\end{equation}

It should be obvious that a proper choice of the regularizing matrix
$R$ is required to provide a satisfactorily small value of
$\overline{||\bfx - \bar \bfx||^2}$. On the other hand, a desired type
of the solution $\bar \bfx$ cannot be determined or even reliably 
estimated from the results of a QMC simulation alone.
Construction of a regularizing algorithm presumes utilization of
certain \textit{a priori} information about the properties of the
solution.

Generally, use of \textit{a priori} information implies the assumption
that the resulting $\bar{\bfx}$ is not arbitrary, but falls into a certain
class of possible solutions.
For example, one can suppose that the resulting function is smooth,
has a particular order of magnitude, \textit{etc}.
In other words there is an \textit{a priori} distribution of $\bar{\bfx}$ ``localized''
around this class.
We introduce an average over such a distribution and denote the averaging with
$\langle \ldots \rangle$.
It is reasonable to require that the regularizing functional $R$ delivers 
a minimum to the deviation $\overline{||\bfx - \bar \bfx||^2}$ averaged over
the chosen distribution of possible solutions,
\begin{equation} \label{mean_deviation}
    \langle\overline{||\bfx - \bar \bfx||^2}\rangle = \mathrm{Tr}(X
    \langle A\rangle X - 2 X\langle B\rangle) +
    \mathrm{Tr}\langle\bar \bfx \bar \bfx^\dag \rangle = \min_{R}
\end{equation}
The main idea of the proposed method is to solve this variational
problem with respect to $R$ and then to obtain $\bfx$ from
(\ref{X_def}). One specific choice of the class of possible solutions
is discussed in Section IV. However, already at this point it is 
worth noting that all we actually need to know about
the possible solutions is the first and the second moments of 
the \textit{a priori} distribution of $\bfx$,
since only these quantities appear in (\ref{mean_deviation})
(recall expressions (\ref{A_def}), (\ref{B_def}) for $A$ and $B$).

Solving (\ref{mean_deviation}) with respect to $R$ and taking
into account the additional condition $\delta R = \delta
R^\dag$ one obtains an equation for the matrix $X$
(the trivial solution $X = 0$ does not relate to the problem):
\begin{equation}\label{opt_eq}
    \langle A\rangle X + X \langle A\rangle =
    \langle B\rangle + \langle B^\dag\rangle
\end{equation}
 
In this way we formally find a system of $N(N+1)/2$ linear
equations for all elements of the matrix $X$ (given $\langle B\rangle$
is a square matrix of size $N\times N$).
However, there is a more efficient way to solve equation (\ref{opt_eq}).
 
Let us denote eigenvalues of the matrix $\langle A\rangle$ with 
$\lambda_i$ and perform an unitary transformation to the eigenbasis 
of $\langle A\rangle$. In this basis a solution of equation
(\ref{opt_eq}) can be expressed explicitly (primes denote matrix elements
in the eigenbasis):
\begin{equation}\label{opt_eq_sol}
    {X_{ij}' = \frac{B_{ij}' + (B_{ji}^*)'}{\lambda_i + \lambda_j}, \quad
    1 \leq i,j \leq N}
\end{equation}
 
The change of the basis allows us to organize numerical solving of system
(\ref{opt_eq}) in an efficient way. A diagonalization procedure for
the matrix $\langle A\rangle$ is quite stable because $\langle A\rangle$
is Hermitian. The advantage of this approach in comparison with direct solving of
(\ref{opt_eq}) is that instead of solving a system of $N(N+1)/2$
equations ($\propto N^6$ operations) it is enough to diagonalize
a $N \times N$ matrix ($\propto N^3$ operations).
 
We should note that formula (\ref{opt_eq_sol}) unambiguously
gives finite expressions for all elements of $X$, since its
denominator is positive. Indeed, the matrix $M^\dag M$ is nonnegative-definite.
The matrix $\langle\bar \bfx \bar \bfx^\dag\rangle$ is a correlation matrix, which is also
essentially positive-definite. Finally, the matrix $K_{\bfy}$ is
also positive-definite provided the vector $\bfy$ is defined with nonzero
inaccuracy (all components have a nonzero dispersion). In this way
we ensure that the matrix $\langle A\rangle$ is positive-definite
and therefore all its eigenvalues are positive.

Nevertheless, a practical implementation of the method faces
certain difficulties, because the numerical solution suffers from
round-off errors during the diagonalization process.
A singular value decomposition procedure does not help much.
To get rid of these round-off errors, we have switched to performing
all calculations with a large number of decimal digits using the MPFR library
(Multiple-Precision Floating-point computations with correct Rounding) \cite{MPFR}.

\section{Choice of the representation. Correlation matrix}
 
Now we apply abstract results of the previous section directly to the
analytic continuation of a function $F(\omega)$ from the imaginary
axis to the real one, assuming $F(\omega)$  to be analytic in the
upper half-plane. A QMC simulation gives values of the function $F$ at
Matsubara frequencies $i \omega_k$ on the imaginary axis:

\begin{equation}
    F_k \equiv F(i\omega_k), \quad k=\overline{1,K}
\end{equation}
 
Let $\omega_0$ be a typical energy scale of the problem (in other words, a natural energy unit).
We introduce a conformal mapping of the upper frequency half-plane to the interior of a
circle of an unitary radius with the center at the origin:
\begin{equation}\label{conformal}
    \omega \leftrightarrow z: \quad z = \frac{\omega - i\omega_0}{\omega +
    i\omega_0} , \quad \omega = i\omega_0\frac{1+z}{1-z}
\end{equation}

\begin{figure}
    \begin{center} 
    \includegraphics[width=0.5\columnwidth]{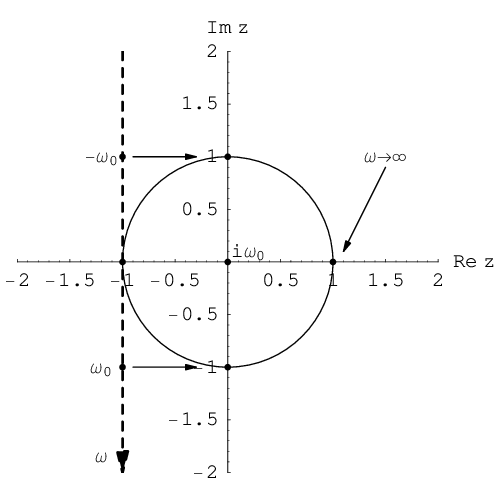}
    \caption{Schematic view of the conformal mapping of the upper frequency
    half-plane to a circle.} \label{fig:conformal}
    \end{center}
\end{figure}
 
The mapping is illustrated in \fref{fig:conformal}. In this
case all imaginary frequencies are univocally mapped onto the segment
$z\in[-1;1]$; all real frequencies correspond to a circle with
a radius 1.
 
If $F$ is analytic in the upper half-plane as a function of a complex
frequency, it has the same property in the unit circle in the
$z$-plane and thus can be expanded in a Taylor series around the point $z=0$:
\begin{equation}\label{F_expansion}
    F(\omega(z)) = \sum_{n=0}^\infty f_nz^n, \quad f_n = \frac{1}{2\pi
    i}\oint\limits_{|z|=1}\frac{F(\omega(z))}{z^{n+1}}dz
\end{equation}

Assuming that the function $F(\omega(z))$ is smooth, we can then take into
account a finite number of terms in this expansion. A proper
number $N$ of terms to be kept can be estimated from test runs of
the algorithm for a fixed model function $F$. When expansion
coefficients $f_n$ are determined, we can sum up the series
at any point of the circle $|z|=1$ and restore values of
$F(\omega)$ on the real axis in this way.
 
In this representation the matrix $M$ from equation (\ref{first_eq})
is a $K\times N$ matrix with elements given by the formula

\begin{equation} M_{kn} =
    \left(\frac{\omega_k-\omega_0}{\omega_k+\omega_0}\right)^n
\end{equation}
 
Knowledge of some \textit{a priori} information about the expected
solutions $F(\omega\in\mathbb{R})$ allows us to choose a correlation matrix $\langle
F(\omega)F^*(\omega')\rangle\mid_{\omega,\omega'\in\mathbb{R}}$
($\langle \bar\bfx \bar\bfx^\dagger\rangle$ in terms of the previous section).
This correlation matrix can also be expanded into a double Taylor
series at zero point, similarly to (\ref{F_expansion}):
\begin{equation}\label{FF_expansion}
    \langle F(\omega)F^*(\omega')\rangle\mid_{\omega,\omega'\in\mathbb{R}}
    \approx \sum_{n,n'=0}^{N-1} \langle f_n f_{n'}^*\rangle z^n
    (z')^{-n'}
\end{equation}
 
\begin{equation}\label{ff_expansion}
    \langle f_n f_{n'}^*\rangle = \frac{1}{(2\pi
    i)^2}\oint\limits_{|z|,|z'|=1} \langle
    F(\omega(z))F^*(\omega(z'))\rangle z^{-(n+1)}z'^{(n'-1)}dz dz'
\end{equation}
 
Practically we have found that $N\approx 30-50$ terms are enough
for any calculation. The value of $\omega_0$ is basically unimportant;
it is sufficient to just take $\omega_0$ of the right order of magnitude.
 
\section{Correlation matrix of Lorentzian peaks}

An explicit form of the correlator $\langle F(\omega)
F^*(\omega')\rangle$ leading to satisfactory results can be
obtained from a simple model describing the \textit{a priori} distribution of
possible solutions $F(\omega)$.
 
Let us assume that the function $F(\omega)$ is a superposition of several
Lorentzian peaks having the same half-width $\gamma$:
\begin{equation}
    F(\omega) = \sum_{j=1}^J \frac{Z_j}{\omega-\Omega_j+i\gamma},
\end{equation}
where positions of the peaks $\Omega_j$ and their residues $Z_i$
are random quantities with known statistical distributions.
\begin{equation}\label{FF}
    \langle F(\omega_1)F^*(\omega_2)\rangle = \sum_{j,j'=1}^J
    \left\langle
    \frac{Z_j}{\omega_1-\Omega_j+i\gamma}\frac{Z_{j'}}{\omega_2-\Omega_{j'}-i\gamma}\right\rangle_{Z,\Omega}
\end{equation}
(averaging $\langle\ldots\rangle$ here is in the sense introduced in Section II).
We assume that all $Z_j$ are equally distributed and uncorrelated with each other,
so there are only two model parameters $\langle Z^2\rangle$ and $\langle Z \rangle$ needed to 
completely describe the residues in the correlation function under consideration:
 
\begin{eqnarray}\label{FF_final}
    \fl\langle F(\omega_1)F^*(\omega_2)\rangle = \langle Z^2
    \rangle\sum_{j=1}^J \left\langle
    \frac{1}{\omega_1-\Omega_j+i\gamma}\frac{1}{\omega_2-\Omega_j-i\gamma}\right\rangle_\Omega
     +\\+ \langle Z \rangle^2 \sum_{j\neq j'}^J \left\langle
\frac{1}{\omega_1-\Omega_j+i\gamma}\frac{1}{\omega_2-\Omega_{j'}-i\gamma}\right\rangle_\Omega
\end{eqnarray}

If we don't have any information about the sign of $\Re F(\omega)$ and $\Im F(\omega)$,
we should put $\langle Z \rangle$ to zero. Otherwise, if such information is available
(as in the case of a Green's function), we can set $\langle Z \rangle$
to a nonzero value. Also $\langle Z \rangle$ and $\langle Z^2
\rangle$ can be used when there is a normalization rule for $F$
which is, in fact, a relation between $Z_j$.
 
Finally we assume, that the pole positions $\Omega_j$ are distributed
independently according to a certain model distribution, for example the
Lorentzian distribution:
\begin{equation}\label{Lorentzian_dist}
    P(\Omega_j) = \frac{1}{\pi\Omega_M}\frac{1}{1+(\Omega_j/\Omega_M)^2}
\end{equation}
where $\Omega_M$ is an estimated total spectrum width. For example,
for the half-filled Hubbard model, discussed in the next section,
a good guess for this quantity is the high-frequency boundary of a Hubbard band.
 
Combination of formulae (\ref{ff_expansion}), (\ref{FF_final}) and (\ref{Lorentzian_dist})
yields a result suitable for practical calculations:
\begin{eqnarray*}
    \fl\langle f_n f_{n'}^*\rangle = J \langle Z^2 \rangle
        \int_{-\infty}^{+\infty} P(\Omega) I(n,\gamma,\Omega)
        I^*(n',\gamma,\Omega)\ d\Omega + \\ +
        J(J-1)\langle Z \rangle^2 \int_{-\infty}^{+\infty} P(\Omega_1)
        I(n,\gamma,\Omega_1)\ d\Omega_1 \int_{-\infty}^{+\infty} P(\Omega_2)
        I(n',\gamma,\Omega_2)\ d\Omega_2
\end{eqnarray*}

\[
I(n,\gamma,\Omega) \equiv \left\{
    \begin{array}{ll}
        \frac{1}{i\gamma -\Omega + i\omega_0}, & n = 0 \\
        -2i\omega_0 \frac{(i\gamma - \Omega - i\omega_0)^{n-1}}{(i\gamma -
        \Omega + i\omega_0)^{n+1}}, & n \neq 0
    \end{array}
\right.
\]
 
Although these integrals could be done analytically for the
particular Lorentzian form of $P(\Omega)$, the answer is very
complicated and has significant computational complexity. It is
therefore much more practical to perform the integration
numerically for every pair of $n$ and $n'$.
 
\section{Hubbard-Mott transition: practical calculation of DOS.}

We have applied the present method to recover the local density of
states of the single band half-filled Hubbard model on a Bethe lattice.
We were aiming at recovering the local DOS as the Hubbard $U$ was changing
and the system was undergoing a Hubbard-Mott transition.

A lot of effort has been done to understand the physics of the Hubbard
model and the most significant progress has been achieved in the case of
a lattice with an infinite coordination, since the Dynamical Mean Field
Theory (DMFT) delivers an exact solution in this limit \cite{RevModPhys.68.13}.
Nevertheless, some questions are still unanswered even in this case. One of them is
about the nature of a fine structure of the Hubbard bands.

There are high-resolution results for the local DOS by Karski \textit{et al} \cite{HubbardFeatures}, 
which were achieved using a standard DMFT procedure at the zero temperature and the D-DMRG impurity
solver. In that calculation an existence of narrow resonances near the inner bounds
of the Hubbard bands in the vicinity of a Mott transition has been reliably established.
Search for similar features at a low but finite temperature by means of a QMC solver
and an analytical continuation procedure would be a good benchmark to compare the proposed
method and the MaxEnt.

In the DMFT approach the single band half-filled Hubbard model is reduced
to the effective impurity problem:
\begin{eqnarray*}
    \fl S_{eff} = -\iint\limits_0^\beta d\tau d\tau' \sum_\sigma
        \bar c_\sigma(\tau)(-\partial_{\tau'} - \Delta(\tau-\tau')) c_\sigma(\tau') +
        U\int_0^\beta d\tau 
            \left(n_\uparrow(\tau) - \frac{1}{2}\right)
            \left(n_\downarrow(\tau) - \frac{1}{2}\right)
        \\
    G(\tau - \tau') =
	\frac{\int \mathcal{D}[\bar c, c] \bar c(\tau') c(\tau) e^{-S_{eff}}}
        {\int \mathcal{D}[\bar c, c] e^{-S_{eff}}},
\end{eqnarray*}

where the hybridization function $\Delta(\tau-\tau')$ is to be determined
self-consistently. For the simplest case of the Bethe lattice the self-consistentcy
condition reads: 
\begin{equation}
    \Delta(\tau-\tau') = t^2 G(\tau-\tau')
\end{equation}

We have performed a number of DMFT runs using the TRIQS \cite{TRIQS,ALPS13,ALPS20} toolkit and its 
implementation of the hybridization expansion algorithm \cite{Werner,ctqmcrmp}. To increase the quality of
the imaginary-time (Matsubara frequency) data we were accumulating the data in the orthogonal 
polynomial representation \cite{polynomial}.

In all the calculations the hopping constant was $t=0.5$ (that corresponds to the bare half-bandwidth $D=1$)
and the inverse temperature was set to $\beta=100$. We considered 3 values of $U$: $U=2.0$ (correlated metal),
$U=2.4$ (vicinity of the Mott phase transition) and $U=3.0$ (Mott insulator with a well-developed gap).

As mentioned above, our algorithm can be applied not only to a thermal Green's function, but also
to a self-energy function or to any other dynamical quantity obeying the same analyticity conditions.
In fact, we used two different paths for the insulating case of large $U$ and for the case when
a quasiparticle peak was present:
\begin{enumerate}
  \item For $U=2.0$ and $U=2.4$ we calculated $A(\omega)$ according to the following scheme:
    \[
	G(i\omega) \mapsto
	\Sigma(i\omega) \mapsto \textrm{(stochastic regularization)} \mapsto \Sigma(\omega) 
	\mapsto G^R(\omega) \mapsto A(\omega),
    \]
    where $\Sigma(i\omega)$ is an auxiliary self-energy defined as $\Sigma(i\omega) = i\omega - G^{-1}(i\omega)$
    (and identically for real frequencies). This quantity has better asymptotic properties than $G(i\omega)$
    (decays faster than $1/i\omega$) and therefore requires a smaller number of $z$-terms in (\ref{F_expansion}) to be
    represented with a given accuracy. Obviously, this positively affects stability of the algorithm.

  \item For $U=3.0$ we used a more straightforward way to do the continuation:
  \[
      G(i\omega) \mapsto \textrm{(stochastic regularization)} \mapsto G^R(\omega) \mapsto A(\omega)
  \]
  In spite of the arguments of the previous paragraph, we have to use an ordinary thermal Green's function,
  since $\Sigma(\omega)$ is not a proper function anymore. It turns to infinity at any point of the insulating
  gap, so it is impossible to construct a valid $z$-representation of it.
\end{enumerate}

\begin{figure}
    \begin{center}
	\includegraphics[scale=0.7]{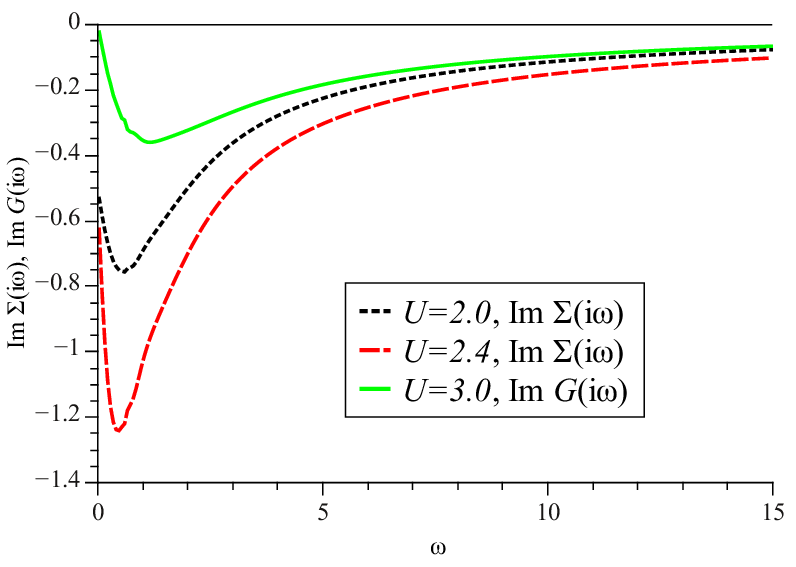}
	\caption{Input data for the stochastic regularization algorithm: 
	$\Im\Sigma(i\omega)$ for $U=2.0$, $U=2.4$ and $\Im G(i\omega)$ for $U=3.0$;
	$t = 0.5$, $\beta = 100$.
	Real parts of the functions are vanishing.}\label{fig:input}
    \end{center}
\end{figure}

The input data for the stochastic regularization algorithm are plotted in \fref{fig:input}.

One has to adjust some parameters of the algorithm to obtain the most plausible DOS curve.
Of course, this introduces a dependence on a ``human factor'', so the values mentioned below are 
only more or less optimal. But on the other hand, as we will show below, the qualitative features
of a spectrum are quite persistent even under wide variations of the adjustable parameters.

To proceed, $K_{\bfy}$ (\ref{K_def}) was chosen to be diagonal with all elements equal to $K_0 = 10^{-5}$,
what means that values $G(i\omega)$ (or $\Sigma(i\omega)$) are statistically independent for different Matsubara
frequencies and distributed with the dispersion $K_0$.

We used the correlator $\langle F(\omega)F^*(\omega')\rangle$ described in Section IV with $\gamma=0.1$ and
$\Omega_M = (U + D)/2$ (the outer boundary of the upper Hubbard band). Other parameters of the correlator were
$J=2, \langle Z \rangle =0, \langle Z^2 \rangle =1$ and they did not seem to affect the result considerably.
In a $z$-expansion we took 40 first terms and any furter increase of this number did not change pictures at all.

\begin{figure}
    \begin{center}
	\subfigure[]{
	    \includegraphics[scale=0.6]{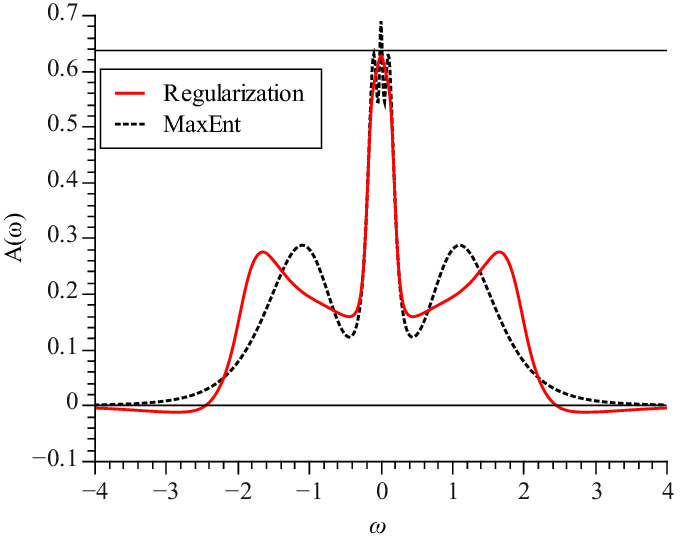}\label{fig:dos20}
	}
	\subfigure[]{
	    \includegraphics[scale=0.6]{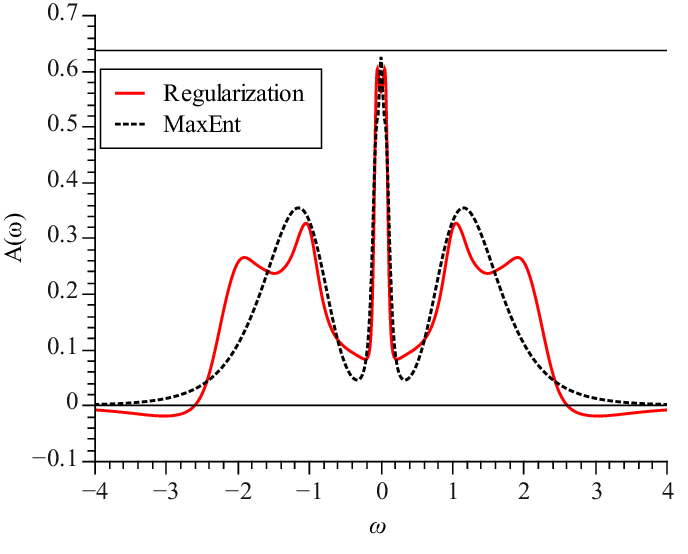}\label{fig:dos24}
	}
	\\
	\subfigure[]{
	    \includegraphics[scale=0.6]{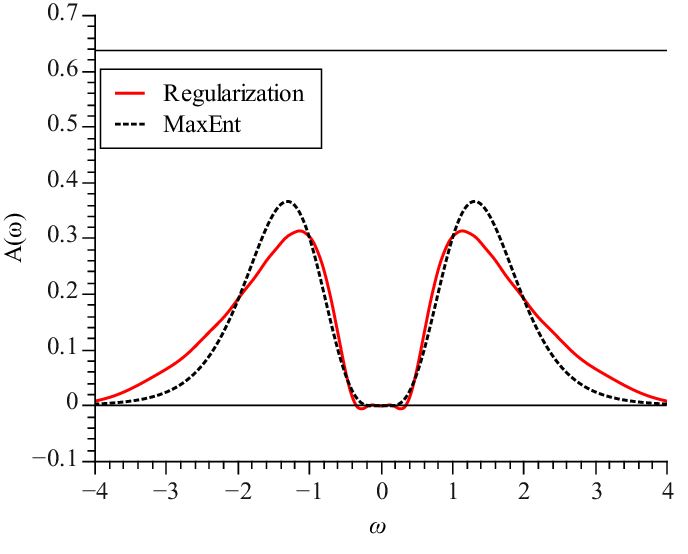}\label{fig:dos30}
	}
	\caption{Reconstructed densities of states for a) $U=2.0$, b) $U=2.4$
	and c) $U=3.0$. Results of the stochastic regularization approach are
	represented by solid lines. MaxEnt predictions are shown with dots. A
	thin horizontal line at $2/\pi$ shows the value of $A(0)$ dictated by the
	Luttinger sum rule.}\label{fig:dos}
    \end{center}
\end{figure}

Resulting local densities of states are depicted in \fref{fig:dos}. For comparison,
we also present results produced by a relatively simple MaxEnt program \cite{PhysRevB.57.10287}
for the same QMC data. A dataset for the regularization program contained values for 1000 first
Matsubara frequencies, while the MaxEnt program worked with 256 (maximally allowed) imaginary time
slices of $G(\tau)$.

We would like to emphasize the presence of additional spectral features (``shoulders'') around $\pm1$
in the \fref{fig:dos24}, which are caught by the present method, but completely overlooked by MaxEnt.
They are reminiscent to what was seen by Karski \textit{et al} \cite{HubbardFeatures} in a zero-temperature
DMFT(D-DMRG) study. On the other hand, there are regions where the stochastic regularization yields
nonphysical negative spectral weight. This problem is a direct consequence of the linearity of our
method: we do not explicitly stipulate positivity. However, the violations of the positivity are not
vast and can be minimized by a proper choice of adjustable parameters.

\begin{figure}
    \begin{minipage}[t]{0.5\linewidth}
	\begin{center}
	    \includegraphics[scale=0.7]{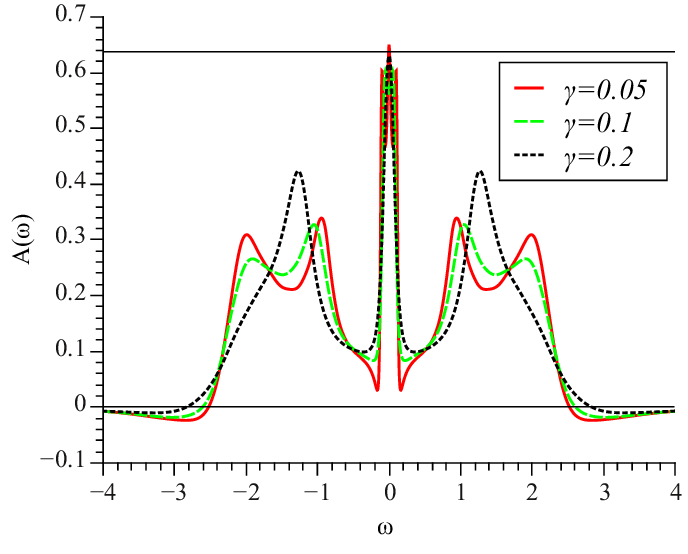}
	    \caption{Reconstructed densities of states for $U=2.4$, $\Omega_M=1.7$, $K_0=10^{-5}$
		    and different values of $\gamma$. The value $\gamma=0.2$ is overestimated and makes
		    two peaks in the Hubbard band merge.}
	    \label{fig:gamma}
	\end{center}
    \end{minipage}
    \begin{minipage}[t]{0.5\linewidth}
	\begin{center}
	    \includegraphics[scale=0.7]{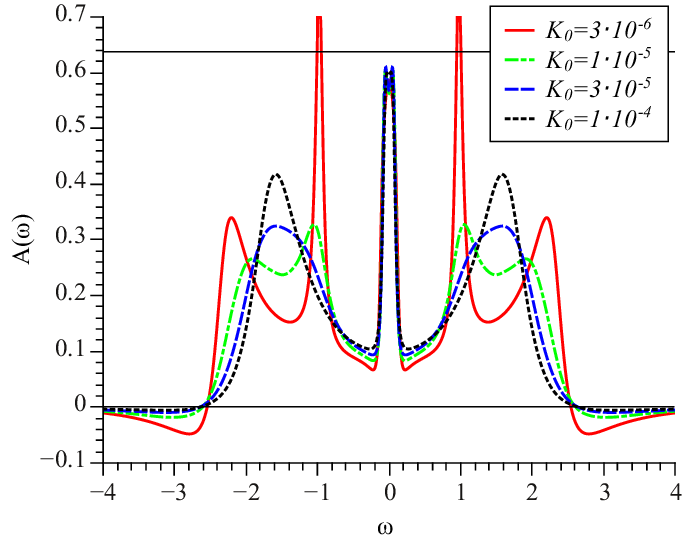}
	    \caption{Reconstructed densities of states for $U=2.4$, $\Omega_M=1.7$, $\gamma=0.1$
		    and different values of $K_0$. With a too large value ($K_0\ge 3\cdot 10^{-5}$) of
		    the estimated noise dispersion, the two-peak structure of the Hubbard bands is
		    lost. For the opposite case of a too small value ($K_0=3\cdot 10^{-6}$) the structure
		    is exaggerated.}
	    \label{fig:K0}
	\end{center}
    \end{minipage}
\end{figure}

Finally, we would like to demonstrate that a DOS produced by our method is indeed affected
by a choice of the parameters, but its qualitative characteristics are quite persistent.
Basically, there are 3 sensible values to be adjusted: $K_0$, $\Omega_M$ and $\gamma$.
$\Omega_M$ defines boundaries of a spectrum and in most situations may be guessed reliably.
$K_0$ tells the algorithm about a desirable level of trust to an input data set. An overestimated
value of $K_0$ may lead to a loss of significant spectral features, while a too small value
binds the result to the input data (and its noise) too tightly and gives just garbage. $\gamma$ 
plays a role of a broadening of spectral features, or in other words controls the amount
of oscillations in a resulting function $F(\omega)$. If the broadening is too large, thin spectral
peaks may be lost again. Sample dependencies of $A(\omega)$
on $\gamma$ and $K_0$ are shown in \fref{fig:gamma} and \fref{fig:K0} respectively.

\section{Conclusion}

To conclude, we have developed a new regularization approach (stochastic regularization)
to the problem of analytic continuation of numerical data from the Matsubara frequencies 
to the real axis. The method is the best possible regularization in sense of
(\ref{mean_deviation}), so that we produce an optimal regularization matrix given
a class of possible solutions and error bars of the input data.

We have applied the proposed method to recover local densities of states of
the single-band Hubbard model on a Bethe lattice at a low temperature. In
contrast to a MaxEnt program, our algorithm was able to resolve
peaks near the inner edges of the Hubbard bands at a value of $U$ where
a Hubbard-Mott transition occurs. Such structures were seen before in a
zero-temperature DMFT(D-DMRG) calculation by Karski \textit{et al}.

Further development should benefit from an important property
of the optimal regularization approach, its linearity with respect to the input data.
It makes possible the continuation of a self-energy or of off-diagonal Green's
function components. In principle, we could follow MaxEnt paradigm
and put additional constrains like the requirement $A(\omega)\ge
0$. However this would make our scheme nonlinear and seriously
reduce its flexibility.

There are also unfavourable consequences of the linearity of our approach: unphysical regions
of negative DOS can occur for certain input data. Another problem of the scheme is related
to its free parameters. In comparison with MaxEnt, more attention is needed from a user to
obtain physical results.

The proposed method probably can not serve as a complete replacement
for MaxEnt due to the mentioned problems, but we believe that it can be helpful in qualitative
resolving of more subtle spectral features which are irreproducible by MaxEnt.

\ack
Authors are thankful to A.I. Lichtenstein, M.I. Katsnelson, S. Biermann and
J. Mravlje for useful discussions, and to P. Crompton for reading the text. The
work was supported by ``Dynasty'' foundation, NWO grant 047.016.005 and RFBR grant
11-02-01443-a. It was also supported by the French ANR under project
CorrelMat and IDRIS/GENCI under project 20111393.

\section*{References}
\bibliographystyle{iopart-num}
\bibliography{regularization}

\providecommand{\newblock}{}
\begin{thebibliography}{10}
\expandafter\ifx\csname url\endcsname\relax
  \def\url#1{{\tt #1}}\fi
\expandafter\ifx\csname urlprefix\endcsname\relax\def\urlprefix{URL }\fi
\providecommand{\eprint}[2][]{\url{#2}}

\bibitem{RevModPhys.68.13}
Georges A, Kotliar G, Krauth W and Rozenberg M~J 1996 {\em Rev. Mod. Phys.\/}
  {\bf 68} 13

\bibitem{kotliar:865}
Kotliar G, Savrasov S~Y, Haule K, Oudovenko V~S, Parcollet O and Marianetti C~A
  2006 {\em Rev. Mod. Phys.\/} {\bf 78} 865--951

\bibitem{maier:1027}
Maier T, Jarrell M, Pruschke T and Hettler M~H 2005 {\em Rev. Mod. Phys.\/}
  {\bf 77} 1027--1080

\bibitem{PhysRevLett.56.2521}
Hirsch J~E and Fye R~M 1986 {\em Phys. Rev. Lett.\/} {\bf 56} 2521--2524

\bibitem{PhysRevB.40.4780}
Fye R~M and Hirsch J~E 1989 {\em Phys. Rev. B\/} {\bf 40} 4780--4796

\bibitem{thirumalai:5029}
Thirumalai D and Berne B~J 1983 {\em The Journal of Chemical Physics\/} {\bf
  79} 5029--5033

\bibitem{PhysRevLett.55.1204}
Sch\"uttler H~B and Scalapino D~J 1985 {\em Phys. Rev. Lett.\/} {\bf 55}
  1204--1207

\bibitem{PhysRevLett.63.2504}
Jarrell M and Biham O 1989 {\em Phys. Rev. Lett.\/} {\bf 63} 2504--2507

\bibitem{MaxEnt_Rev}
Jarrell M and Gubernatis J~E 1996 {\em Physics Reports\/} {\bf 269} 133 -- 195
  ISSN 0370-1573

\bibitem{PhysRevE.81.056701}
Fuchs S, Pruschke T and Jarrell M 2010 {\em Phys. Rev. E\/} {\bf 81} 056701

\bibitem{PhysRevB.62.6317}
Mishchenko A~S, Prokof'ev N~V, Sakamoto A and Svistunov B~V 2000 {\em Phys.
  Rev. B\/} {\bf 62} 6317--6336

\bibitem{Tikhonov}
Tikhonov A and Arsenin V 1977 {\em Solutions of ill-posed problems\/} (VH
  Winston)

\bibitem{MPFR}
http://www.mpfr.org/

\bibitem{HubbardFeatures}
Karski M, Raas C and Uhrig G 2005 {\em Phys. Rev. B\/} {\bf 72} 113110

\bibitem{TRIQS}
Ferrero M and Parcollet O {TRIQS, A toolkit for Research in Interacting Quantum
  Systems} \textit{To be published}

\bibitem{ALPS13}
{Albuquerque A F \textit{et al}} 2007 {\em Journal of Magnetism and Magnetic
  Materials\/} {\bf 310} 1187 -- 1193

\bibitem{ALPS20}
{Bauer B \textit{et al}} 2011 {\em Journal of Statistical Mechanics: Theory and
  Experiment\/} {\bf 2011} P05001

\bibitem{Werner}
Werner P, Comanac A, de'Medici L, Troyer M and Millis A~J 2006 {\em Phys. Rev.
  Lett.\/} {\bf 97} 076405

\bibitem{ctqmcrmp}
Gull E, Millis A~J, Lichtenstein A~I, Rubtsov A~N, Troyer M and Werner P 2011
  {\em Rev. Mod. Phys.\/} {\bf 83} 349--404

\bibitem{polynomial}
Boehnke L, Hafermann H, Ferrero M, Lechermann F and Parcollet O 2011 {\em ArXiv
  e-prints\/} (\textit{Preprint} \eprint{1104.3215})

\bibitem{PhysRevB.57.10287}
Sandvik A~W 1998 {\em Phys. Rev. B\/} {\bf 57} 10287--10290

\end{thebibliography}

\end{document}